\documentclass[preprint,prc,aps,EPSf]{revtex4}
\usepackage{graphicx}
  \def\lambdaBp{\lambda_{n^+}}
  \def\lambdaBm{\lambda_{n^-}}
  \def\mBp{m_+^n}
  \def\mBm{m_-^n}
  \def\mBpm{m_\pm^n}
  \def\g{{\gamma}}

  \def\ss{\langle\bar ss\rangle} 
  \def\qq{\langle\bar qq\rangle} 
  \def\sGs{\langle g\bar s\sigma_{\mu\nu}G^{\mu\nu}s\rangle} 
  \def\qGq{\langle g\bar q\sigma_{\mu\nu}G^{\mu\nu}q\rangle} 
  \def\GG{\langle{\alpha_s\over\pi}G_{\mu\nu}G^{\mu\nu}\rangle}
  \def\<>#1{\langle#1\rangle} 
  \def\><{\rangle\langle} 
  \def\vec#1{\mbox{\boldmath$#1$}}

\begin{document}

\preprint{KEK-TH-1105}
\title{Positive and negative-parity flavor-octet baryons in coupled QCD sum rules}
\author{Yoshihiko Kondo\footnote{kondo@kokugakuin.ac.jp}}
\address{Kokugakuin University, Higashi, Shibuya, Tokyo 150-8440, Japan}
\author{Osamu Morimatsu\footnote{osamu.morimatsu@kek.jp}}
\address{Institute of Particle and Nuclear Studies, High Energy Accelerator Research Organization, 1-1, Ooho, Tsukuba, Ibaraki, 305-0801, Japan}
\author{Tetsuo Nishikawa\footnote{nishi@th.phys.titech.ac.jp}}
\address{Department of Physics, Tokyo Institute of Technology, 2-12-1, Oh-Okayama, Meguro, Tokyo 152-8551, Japan}
\author{Yoshiko Kanada-En'yo \footnote{yenyo@yukawa.kyoto-u.ac.jp}}
\address{Yukawa Institute for Theoretical Physics, Kyoto University, Kyoto 606-8502, Japan}

\begin{abstract}
We apply the method of the QCD sum rule, in which positive- and negative-parity baryons couple with each other, to the flavor-octet hyperons and investigate the parity splittings.
We also reexamine the nucleon in the method, which was studied in our previous paper, by carefully choosing the Borel weight.
Both in the nucleon and hyperon channels the obtained sum rules turn out to have a very good Borel stability and also have a Borel window, an energy region in which the OPE converges and the pole contribution dominates over the continuum contribution.
The predicted masses of the positive- and negative-parity baryons reproduce the experimental ones fairly well in the $\Lambda$ and $\Sigma$ channels, if we assign the $\Lambda(1670)$ and the $\Sigma(1620)$ to the parity partners of the $\Lambda$ and the $\Sigma$, respectively.
This implies that the $\Lambda(1405)$ is not the party partner of the $\Lambda$ and may be a flavor-singlet or exotic state.
In the $\Xi$ channel, the sum rule predicts the mass of the negative-parity state to be about 1.8~GeV, which leads to two possibilities; one is that the observed state with the closest mass, $\Xi(1690)$, is the parity partner and the other is that the parity partner is not yet found but exists around 1.8~GeV.
\end{abstract}

\maketitle

\newpage
\section{Introduction}

One of the unsolved problems in the physics of hadron spectroscopy is the negative-parity baryon.
Experimentally, negative-parity baryons in hyperon channels have not been sufficiently studied.
There are little data in the $\Xi$ channel and spin-parity quantum numbers of many resonances are not determined in the $ \Lambda$ and $\Sigma$ channels~\cite{PDG}.
Theoretically, chiral symmetry is a key issue in QCD together with confinement and therefore it is important to understand how it manifests itself in the baryon sector.
Positive and negative-parity baryons have different masses in the broken (Goldstone) phase of chiral symmetry while they should be degenerate in the unbroken (Wigner) phase.
Thus, mass differences of positive and negative-parity baryons should be understood as a consequence of the spontaneous breakdown of chiral symmetry.
However, this has not been achieved yet.
In constituent quark models, negative-parity baryons are usually regarded as states in which one of three constituent quarks is excited to a p-wave orbit, and the relation to the spontaneous breakdown of chiral symmetry is unclear~\cite{Isgur,Glozman,Furuichi}.
The lattice QCD, which is the most legitimate nonperturbative method, has difficulties in applying to negative-parity baryons because the negative-parity baryon is located in the continuum of the positive-parity baryon and the meson~\cite{Leinweber,Nemoto}.
In the QCD sum rule approach, the properties of hadrons, such as the mass or the decay constant, are related to the vacuum condensates, i.e. vacuum expectation values of quark-gluon composite operators~\cite{SVZ}.
Since the spontaneous breakdown of chiral symmetry is associated with condensates of chiral-odd operators, the method of the QCD sum rule seems very suitable for the purpose of understanding the parity splitting of baryons from the spontaneous breakdown of chiral symmetry~\cite{chungII,jido,lee}.
Though the method of the QCD sum has the same difficulty as does the lattice QCD, there have been some attempts to overcome the difficulty and apply the method to negative-parity baryons.
Still, their sum rules are not very satisfactory in a sense that they do not have a sufficient Borel stability.

Recently, we proposed a new approach of the QCD sum rule for the baryon in which positive- and negative-parity baryons couple with each other~\cite{OUR}.
Then, we applied the method to the nucleon channel and investigated the parity splitting of the nucleon.
We found that the Borel stability of the sum rule has been drastically improved not only for the positive-parity nucleon but also for the negative-parity nucleon.
The obtained results showed that the parity splitting of baryons can be studied in the framework of the QCD sum rule once the coupling of
positive- and negative-parity states is appropriately taken into account.
The following point, however, remained unclarified in the previous work~\cite{OUR}.
For the positive-parity nucleon we found a sum rule with an energy region in which the operator product expansion (OPE) reasonably converges and contribution from the continuum is smaller than the pole contribution, while such a sum rule cannot be found for the negative-parity nucleon.

The purpose of the present paper is two-fold.
Our first aim is to reexamine the nucleon channel in the coupled QCD sum rule more carefully.
We would like to see if it is really impossible to construct sum rules for the negative-parity nucleon which has convergence of the OPE and dominance of the pole over the continuum.
Our second aim is to apply the method of the coupled QCD sum rules to the hyperon channels and investigate the parity splitting of the hyperons.
We would like to see which observed states are the parity partners of the lowest positive-parity hyperons and also study the effect of the SU(3) breaking on negative-parity baryons.
There is a well-known problem in the negative-parity hyperon, the problem of the $\Lambda(1405)$.
In the $\Lambda$ channel the lowest negative-parity state is the $\Lambda(1405)$.
However, many theoretical calculations have failed to reproduce its mass.
This may imply that the parity partner of the $\Lambda$ is not the $\Lambda(1405)$ but a higher resonance.
In fact, the idea has been proposed that $\Lambda(1405)$ is not a flavor-octet state or not even a simple three-quark state.
We would like to clarify if the sum rules conclude that the $\Lambda (1405)$ can be regarded as the parity partner of the $\Lambda$ or not.

\section{Coupled QCD sum rule}

We consider the correlation function of the baryon interpolating field, $\eta$, as 
\begin{eqnarray}
\Pi(p)=-i\int{d^4x}e^{ipx}\langle0|T(\eta(x)\bar\eta(0))|0\rangle.
\end{eqnarray}
The field, $\eta$, couples to positive- and negative-parity resonance states, $|B^n_+\rangle$ and $|B^n_-\rangle$, respectively, as
\begin{eqnarray}
\langle0|\eta(x)|B^n_+(p,s)\rangle&=&\lambdaBp u_{B^n_+}(p,s)e^{-ipx},\cr
\langle0|\eta(x)|B^n_-(p,s)\rangle&=&\lambdaBm \gamma_5u_{B^n_-}(p,s)e^{-ipx},
\end{eqnarray}
where $u_B(p,s)$ is a positive energy solution of the free Dirac equation of the baryon~\cite{chungII}.
The spectral function, $\rho(p_0)\equiv-{\rm Im}\Pi(p_0+i\epsilon)/\pi$, can be expressed as the sum of the positive- and negative-parity baryon states:
\begin{eqnarray}\label{Spectral}
\rho(p_0)&=&
P_+\sum_n\left\{|\lambdaBp|^2\delta(p_0-\mBp)+|\lambdaBm|^2\delta(p_0+\mBm)\right\}
\cr&&+P_-\sum_n\left\{|\lambdaBp|^2\delta(p_0+\mBp)+|\lambdaBm|^2\delta(p_0-\mBm)\right\},
\end{eqnarray}
where $P_\pm=(\g_0\pm1)/2$ and $\mBpm$ is the mass of the $n$-th resonance. 
In Eq.~(\ref{Spectral}) we take the zero-width approximation and the rest frame, $\vec p=0$~\cite{chungII,jido}. 

Now, let us consider the projected spectral function,
\begin{eqnarray}
{\rho_\pm}(p_0)={1\over4}{\rm Tr}\left[P_\pm{\rho}(p_0)\right],\quad
\rho_\pm^{\rm OPE}(p_0)={1\over4}{\rm Tr}\left[P_\pm{\rho}^{\rm OPE}(p_0)\right],
\end{eqnarray}
where ${\rho}^{\rm OPE}$ means the spectral function in the OPE.
The projected spectral function satisfies a relation,
\begin{eqnarray}
{\rho_\pm}(p_0)={\rho_\mp}(-p_0),\quad 
\rho_\pm^{\rm OPE}(p_0)=\rho_\mp^{\rm OPE}(-p_0).
\end{eqnarray}
We approximate each of the positive- and negative-parity contributions in the projected spectral function by the lowest pole plus continuum ansatz.
Namely, we parameterize the projected spectral function as,
\begin{eqnarray}\label{Phen}
\rho_\pm(p_0)&=&|\lambda_\pm|^2\delta(p_0-m_\pm)+|\lambda_\mp|^2\delta(p_0+m_\mp)
+[\theta(p_0-\omega_\pm)+\theta(-p_0-\omega_\mp)]\rho^{\rm OPE}_\pm(p_0),
\end{eqnarray}
where $\omega_+$ and $\omega_-$ denote the effective continuum thresholds for positive- and negative-parity channels, respectively.
One can obtain the QCD sum rule by using the analyticity of the correlation function and replacing the correlation function, $\Pi$, by that in the OPE, $\Pi^{\rm OPE}$, in the deep Euclid region, $p_0^2\rightarrow-\infty$, as
\begin{eqnarray}\label{QSR}
\int_{-\infty}^{\infty}dp_0\rho_\pm^{\rm OPE}(p_0)W(p_0)
&=&\int_{-\infty}^{\infty}dp_0\rho_\pm(p_0)W(p_0).
\end{eqnarray}
Substituting Eq.~(\ref{Phen}) to the right-hand side in Eq.~(\ref{QSR}) and using the Borel weight, $W(p_0)=p_0^n\exp(-p_0^2/M^2)$, we obtain the Borel sum rule,
\begin{eqnarray}\label{BSR}
\Pi_\pm^n(M,\omega_+,\omega_-)&\equiv&
\int_{-\omega_\mp}^{\omega_\pm}dp_0\rho_\pm^{\rm OPE}(p_0)p_0^n\exp(-{p_0^2\over M^2})
\cr&=&{m_\pm}^n|\lambda_\pm|^2\exp(-{{m_\pm}^2\over M^2})+(-m_\mp)^n|\lambda_\mp|^2\exp(-{{m_\mp}^2\over M^2}),
\end{eqnarray}
where the parameter of the weight function, $M$, is called the Borel mass.
The projected correlation function defined by Eq.~(\ref{BSR}) satisfies a relation,
\begin{eqnarray}\label{Relation}
\Pi_\pm^n(M,\omega_+,\omega_-)=(-1)^n\Pi_\mp^n(M,\omega_+,\omega_-).
\end{eqnarray}
It should be noted that in Eq.~(\ref{BSR}) the positive- and negative-parity baryons couple with each other, which provides us with a relation between condensates and phenomenological parameters, $m_\pm$, $\lambda_\pm$ and $\omega_\pm$.
Combining $\Pi_+^n$ ($\Pi_-^n$) defined by Eq.~(\ref{BSR}) with different $n$, we can eliminate $\lambda_\pm$ as
\begin{eqnarray}\label{SRtwo}
m_\pm&=&\Bigg[\sqrt{(\Pi_+^k\Pi_-^{l+2}-\Pi_+^{k+2}\Pi_-^l)^2
 +4(\Pi_+^k\Pi_-^{l+1}+\Pi_+^{k+1}\Pi_-^l)(\Pi_+^{k+1}\Pi_-^{l+2}+\Pi_+^{k+2}\Pi_-^{l+1})}
\cr&&\mp(\Pi_+^k\Pi_-^{l+2}-\Pi_+^{k+2}\Pi_-^l)\Bigg]\left/2(\Pi_+^k\Pi_-^{l+1}+\Pi_+^{k+1}\Pi_-^l)\right.\qquad(l\not=k).
\end{eqnarray}
Eq.~(\ref{SRtwo}) expresses the baryon mass, $m_\pm$, as a function of the Borel mass, $M$, and the effective continuum thresholds, $\omega_\pm$, as well as condensates.
In our previous work~\cite{OUR}, we took the effective continuum thresholds to be the energies of first exited baryons in positive- and negative-parity channels for simplicity and determined the baryon masses by Eq.~(\ref{SRtwo}).
In the present paper, we will determine both the baryon masses and the effective continuum thresholds
by the sum rules, which is in better accordance with the philosophy of the QCD sum rule.
We will explain the details of the procedure in the next section.

\section{Masses of  flavor-octet baryons}

The interpolating fields for the flavor-octet baryons without derivatives~\cite{ioffe,espriu,ioffeII} are given by
\begin{eqnarray}\label{BIF}
\eta_N&=&\epsilon_{abc}[(u_aCd_b)\g_5u_c+t(u_aC\g_5d_b)u_c],
\cr
\eta_\Lambda&=&\sqrt{1\over6}\epsilon_{abc}[(d_a(x)Cs_b(x))\g_5u_c(x)+(s_a(x)Cu_b(x))\g_5d_c(x)-2(u_a(x)Cd_b(x))\g_5s_c(x)
\cr&&+t\{(d_a(x)C\g_5s_b(x))u_c(x)+(s_a(x)C\g_5u_b(x))d_c(x)-2(u_a(x)C\g_5d_b(x))s_c(x)\}],
\cr
\eta_\Sigma&=&\epsilon_{abc}[(u_a(x)Cs_b(x))\g_5u_c(x)+t(u_a(x)C\g_5s_b(x))u_c(x)],
\cr
\eta_\Xi&=&\epsilon_{abc}[(s_a(x)Cu_b(x))\g_5s_c(x)+t(s_a(x)C\g_5u_b(x))s_c(x)],
\end{eqnarray}
where $u$, $d$ and $s$ are field operators of up, down and strange quarks, respectively, $C$ denotes the charge conjugation operator and $a$, $b$ and $c$ are color indices. 
In Eq.~(\ref{BIF}), $t$ can be regarded as the tangent of a mixing angle and is called the mixing parameter in the following.

The OPE expressions of the spectral functions for the flavor-octet baryons are given by
\begin{eqnarray}\label{Nrho}
\rho^N(p_0)&=&\gamma_0\{{5+2t+5t^2\over2^{11}\pi^4}p_0^5[\theta(p_0)-\theta(-p_0)]
\cr&&+{5+2t+5t^2\over2^{10}\pi^2}\GG p_0[\theta(p_0)-\theta(-p_0)]
\cr&&+{7t^2-2t-5\over24}\qq^2\delta(p_0)\}
\cr&&-{7t^2-2t-5\over64\pi^2}p_0^2\qq[\theta(p_0)-\theta(-p_0)]
\cr&&+{3(t^2-1)\over64\pi^2}\qGq[\theta(p_0)-\theta(-p_0)],
\end{eqnarray}
\begin{eqnarray}
\rho^{\Lambda}(p_0)&=&\g_0\Bigg\{{5+2t+5t^2\over2^{11}\pi^4}p_0^5[\theta(p_0)-\theta(-p_0)]
+\Bigg({5+2t+5t^2\over2^{10}\pi^2}\GG
\cr&&+{1+4t-5t^2\over96\pi^2}m_s\qq+{5+2t+5t^2\over128\pi^2}m_s\ss\Bigg)p_0[\theta(p_0)-\theta(-p_0)]
\cr&&+\Bigg(-{13-2t-11t^2\over72}\qq^2-{1+4t-5t^2\over36}\qq\ss
\cr&&-{1+t+t^2\over2^5 3\pi^2}m_s\sGs-{7+4t-11t^2\over2^7 3\pi^2}m_s\qGq\Bigg)\delta(p_0)
\cr&&+{1-t^2\over2^6\pi^2}m_s\qGq{1\over p_0}[\theta(p_0-m_s)-\theta(-p_0-m_s)]
\Bigg\}
\cr&&-{13-2t-11t^2\over2^{9}3\pi^4}m_sp_0^4[\theta(p_0)-\theta(-p_0)]
\cr&&+\Bigg({1+4t-5t^2\over96\pi^2}\qq+{13-2t-11t^2\over192\pi^2}\ss\Bigg)p_0^2[\theta(p_0)-\theta(-p_0)]
\cr&&+\Bigg(-{1-t^2\over2^5\pi^2}\sGs-{1-t^2\over2^6\pi^2}\qGq\Bigg)[\theta(p_0)-\theta(-p_0)],
\end{eqnarray}
\begin{eqnarray}
\rho^{\Sigma}(p_0)&=&\g_0\Bigg\{{5+2t+5t^2\over2^{11}\pi^4}p_0^5[\theta(p_0)-\theta(-p_0)]
+\Bigg({5+2t+5t^2\over2^{10}\pi^2}\GG
\cr&&+3{1-t^2\over32\pi^2}m_s\qq+{5+2t+5t^2\over128\pi^2}m_s\ss\Bigg)p_0[\theta(p_0)-\theta(-p_0)]
\cr&&+\Bigg({1-2t+t^2\over24}\qq^2-{1-t^2\over4}\qq\ss
\cr&&+{1+t+t^2\over2^5 3\pi^2}m_s\sGs-13{1-t^2\over2^7\pi^2}m_s\qGq\Bigg)\delta(p_0)
\cr&&+3{1-t^2\over2^6\pi^2}m_s\qGq{1\over p_0}[\theta(p_0-m_s)-\theta(-p_0-m_s)]
\Bigg\}
\cr&&+{1-2t+t^2\over2^{9}\pi^4}m_sp_0^4[\theta(p_0)-\theta(-p_0)]
\cr&&+\Bigg(3{1-t^2\over32\pi^2}\qq-{1-2t+t^2\over64\pi^2}\ss\Bigg)p_0^2[\theta(p_0)-\theta(-p_0)]
\cr&&-3{1-t^2\over2^6\pi^2}\qGq[\theta(p_0)-\theta(-p_0)]
\end{eqnarray}
and
\begin{eqnarray}
\rho^{\Xi}(p_0)&=&\g_0\Bigg\{{5+2t+5t^2\over2^{11}\pi^4}p_0^5[\theta(p_0)-\theta(-p_0)]
+\Bigg({5+2t+5t^2\over2^{10}\pi^2}\GG
\cr&&+3{1-t^2\over32\pi^2}m_s\qq+3{1+2t+t^2\over64\pi^2}m_s\ss\Bigg)p_0[\theta(p_0)-\theta(-p_0)]
\cr&&+\Bigg({1-2t+t^2\over24}\ss^2-{1-t^2\over4}\qq\ss
\cr&&-{1+10t+t^2\over2^7 3\pi^2}m_s\sGs-5{1-t^2\over2^7\pi^2}m_s\qGq\Bigg)\delta(p_0)
\cr&&+3{1-t^2\over2^6\pi^2}m_s\qGq{1\over p_0}[\theta(p_0-2m_s)-\theta(-p_0-2m_s)]
\Bigg\}
\cr&&-3{1-t^2\over2^{8}\pi^4}m_s p_0^4[\theta(p_0)-\theta(-p_0)]
\cr&&+\Bigg(3{1-t^2\over32\pi^2}\ss-{1-2t+t^2\over64\pi^2}\qq\Bigg)p_0^2[\theta(p_0)-\theta(-p_0)]
\cr&&-3{1-t^2\over2^6\pi^2}\sGs[\theta(p_0)-\theta(-p_0)],
\end{eqnarray}
where we adopted the factorization hypothesis for the four-quark condensate and $\qq\equiv\<>{\bar uu}=\<>{\bar dd}$.
In the gluon condensate, $\GG$, $\alpha_s=g^2/4\pi$ and $G_{\mu\nu}= G_{\mu\nu}^a\lambda^a/2$, where $g$, $G_{\mu\nu}^a$ and $\lambda^a$ denote the strong coupling constant, the gluon field tensor and the usual Gell-Mann SU(3) matrix, respectively.
For the QCD parameters we choose the standard values~\cite{ioffe,RRY,RRYII,BandI,SVVZ,OandP}:
\begin{eqnarray}\label{Parameter}
\qq=-(0.23\;{\rm GeV})^3,\cr
\GG=(0.33\;{\rm GeV})^4,\cr 
m_0^2\equiv{\qGq\over\qq}=0.65\;{\rm GeV}^2,\cr
m_s=0.1\;{\rm GeV},\cr 
\ss=0.8\qq,\cr 
{\sGs\over\ss}=1\;{\rm GeV}^2.
\end{eqnarray}

\begin{table}[t]
\caption{Effective continuum thesholds and mixing parameters.}
\label{TableI}
{\setlength{\tabcolsep}{30pt}
\begin{tabular}{c|ccccc}
\hline\hline
Channel  &$\omega_+$&$\omega_-$&  $t$  \\
         &   (GeV)  &   (GeV)  &       \\ \hline
$N$      &    1.4   &    1.7   &$-0.71$\\
$\Lambda$&    1.5   &    1.8   &$-0.76$\\
$\Sigma$ &    1.5   &    1.8   &$-0.70$\\
$\Xi$    &    1.7   &    2.0   &$-0.78$\\
\hline\hline
\end{tabular}}
\end{table}

Now we calculate the positive- and negative-parity baryon masses from the sum rules~(\ref{SRtwo}).
We determine the phenomenological parameters from the condition that they are least sensitive to the Borel weight.
Namely, we search for the effective continuum thresholds, $\omega_+$ and $\omega_-$, and the mixing parameter, $t$, for which the obtained baryon masses with two choices of the Borel weight, $\{k,l\}=\{0,1\}$ and $\{0,2\}$, become as close to each other as possible and the Borel curves for baryon masses have most stable plateaus.
Table~\ref{TableI} shows the values of the $\omega_+$, $\omega_-$ and $t$ determined as described above.
For this choice of parameters, the Borel stability appears within the range of $M=0.8-1.8$~GeV with $\{k,l\}=\{0,1\}$ and $\{0,2\}$, and the obtained baryon masses are roughly the same, as we will show.

It should be noted that for the sum rule to be reliable the OPE must converge and the pole contribution must dominate over the continuum contribution, which can be realized in a restricted range of the Borel mass if at all possible.
From here on, we call the energy region a Borel window, which satisfies that in the OPE the contribution of the highest order included terms is less than 10 \% of the total sum and that the continuum contribution is less than the pole contribution.
In our previous work~\cite{OUR}, we made an optimal choice of the parameters only from the condition that the sum rule for the positive-parity nucleon has a Borel window.
Then, the sum rule for the negative-parity nucleon had no Borel window because of its large continuum contribution.
The absence of the Borel window for the negative-parity baryon seemed natural since the negative-parity nucleon is above the positive-parity continuum threshold.
Surprisingly, by reexamining the sum rules carefully, we have found sum rules with a Borel window for not only positive but also negative-parity baryons, which we will show in the following.

\begin{figure}
  \begin{center}
\includegraphics[width=9.7cm]{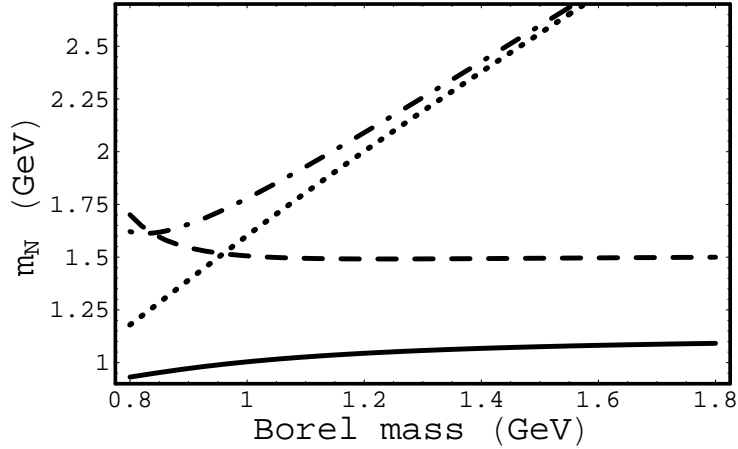}
  \end{center}
\caption{
Nucleon masses as a function of Borel mass for the Borel weight, $\{k,l\}=\{0,1\}$.
The solid and the dashed lines correspond to the positive- and the negative-parity nucleon, respectively.
The dotted (dot-dashed) line shows the positive-parity (negative-parity) nucleon mass without the continuum.\\
}
  \label{figI}
\end{figure}

\begin{figure}
  \begin{center}
\includegraphics[width=9.7cm]{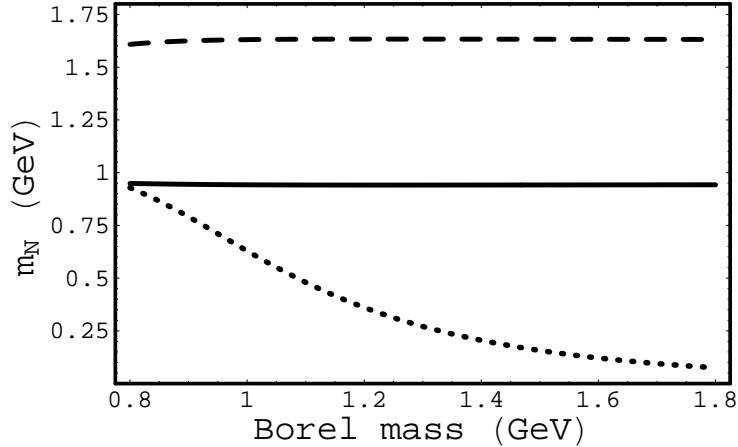}
  \end{center}
\caption{
Nucleon masses as a function of Borel mass for the Borel weight, $\{k,l\}=\{0,2\}$.
The solid and the dashed lines correspond to the positive- and the negative-parity nucleon, respectively.
The dotted line shows the positive-parity nucleon mass without the continuum, while the line of the negative-parity is not displayed because it is outside the range of the figure.
}
  \label{figII}
\end{figure}

We first study the sum rules for the nucleon channel in detail.
Figures~\ref{figI} and \ref{figII} show the Borel curves for the masses of positive- and negative-parity nucleons with $\{k,l\}=\{0,1\}$ and $\{k,l\}=\{0,2\}$, respectively, which also include the curves without continuum contribution for comparison.
The curve without continuum contribution, however, is not shown for the negative-parity nucleon in Fig.~\ref{figII}, because it is outside the range of the figure.
On the one hand, one sees from Fig.~\ref{figI} that the continuum contributes less than 50\% to the positive-parity (negaive-parity) mass when the Borel mass is below $\approx 0.9$ $(1.3)$~GeV and $\{k,l\}=\{0,1\}$.
Simiarly, from Fig.~\ref{figII} the continuum contribution in the positive-parity (negative-parity) mass is less than 50\% when the Borel mass is below $\approx1.1$ ($\sim0.7$)~GeV and $\{k,l\}=\{0,2\}$.
On the other hand, we checked that the highest-order terms, dimension-six terms, in the OPE contribute to the nucleon mass less than 10\% both in the positive- and negative-parity channels when the Borel mass is above 0.8~GeV.
Moreover, the Borel curve has a perfectly stable plateau for the positive-parity nucleon mass in Fig.~\ref{figII} and a reasonably stable plateau for the negative-parity nucleon mass in Fig.~\ref{figI} in the range of $M=1-1.3$~GeV.
From the values of the plateaus we obtain the results of the sum rules for the masses of the positive- and negative-parity nucleons as
\begin{eqnarray}
m_+^N=0.94\;{\rm GeV},\quad m_-^N=1.49\;{\rm GeV}.
\end{eqnarray}
These values reproduce the experimental masses, $m_+^N=0.939\;{\rm GeV}$
and $m_-^N=1.535\;{\rm GeV}$, within 5\%.

\begin{figure}
  \begin{center}
\includegraphics[width=9.7cm]{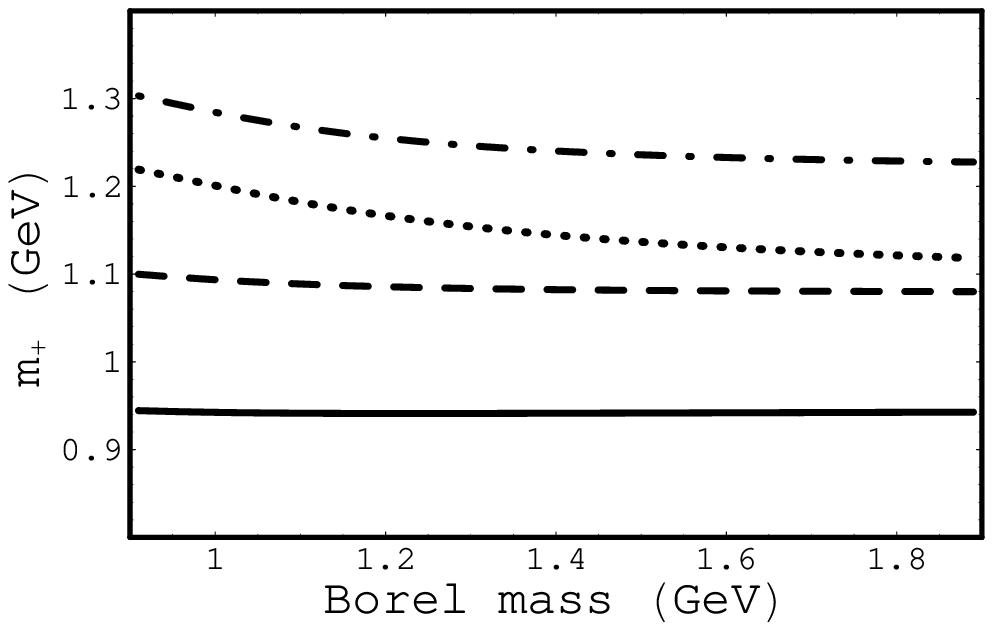}
  \end{center}
\caption{
Positive-parity baryon masses calculated in Eq.(\ref{SRtwo}) with the Borel weight, $\{k,l\}=\{0,2\}$, as a function of Borel mass.
The solid, the dashed, dotted and dot-dashed lines correspond to the nucleon, $\Lambda$, $\Sigma$ and $\Xi$  channels, respectively.
}
  \label{figIII}
\end{figure}

\begin{figure}
  \begin{center}
\includegraphics[width=9.7cm]{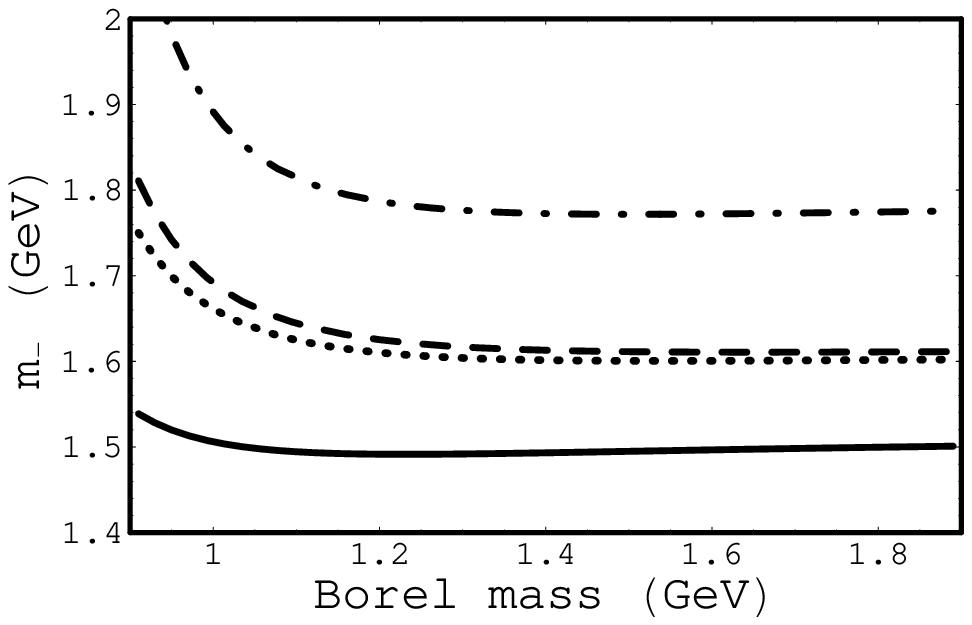}
  \end{center}
\caption{
Negative-parity baryon masses calculated in Eq.(\ref{SRtwo}) with the Borel weight, $\{k,l\}=\{0,1\}$, as a function of Borel mass.
The solid, the dashed, dotted and dot-dashed lines correspond to the nucleon, $\Lambda$, $\Sigma$ and $\Xi$ channels, respectively.
}
  \label{figIV}
\end{figure}

We next move on to the hyperon channels.
Concerning the stability of the sum rules the hyperon channels have tendencies similar to the nucleon channel.
Namely, the Borel weight $\{k,l\}=\{0,2\}$ ($\{k,l\}=\{0,1\}$) gives the most stable sum rule for the positive-parity (negative-parity) hyperons.
Figures~\ref{figIII} and \ref{figIV} show the Borel curves for the masses of positive- and negative-parity flavor-octet baryons, respectively.
In Fig.~\ref{figIII} one sees that the Borel curves calculated by Eq.~(\ref{SRtwo}) with $\{k,l\}=\{0,2\}$ are fairly stable in the whole range of the figure.
These curves turn out to have roughly the same Borel window, $M=0.9-1.2$~GeV. 
From the values in the Borel window we obtain
\begin{eqnarray}
m_+^\Lambda=1.1\;{\rm GeV},\cr
m_+^\Sigma=1.2\;{\rm GeV},\cr
m_+^\Xi=1.3\;{\rm GeV}.
\end{eqnarray}
These values reproduce the experimental masses, $m_+^\Lambda=1.116\;{\rm GeV}$, $m_+^\Sigma=1.193\;{\rm GeV}$ and $m_+^\Xi=1.318\;{\rm GeV}$, fairly well.
From the sum rule for the masses of negative-parity baryons, Eq.~(\ref{SRtwo}) with $\{k,l\}=\{0,1\}$, we find that the Borel window is about $M=0.9-1.4$~GeV. 
In Fig.~\ref{figIV} one sees that the Borel curves are stable in the region of $M>1.2$~GeV.
Taking the values in the region of $M=1.2-1.4$~GeV we obtain
\begin{eqnarray}
m_-^\Lambda=1.6\;{\rm GeV},\cr
m_-^\Sigma=1.6\;{\rm GeV},\cr
m_-^\Xi=1.8\;{\rm GeV}.
\end{eqnarray}
The obtained values of $m_-^\Lambda$ and $m_-^\Sigma$ are close to the Breit-Wigner masses of the $\Lambda(1670)$ and the $\Sigma(1620)$, respectively.
The results of the sum rules suggest that the parity partner of the $\Lambda$ should be assigned to the $\Lambda(1670)$ and that of the $\Sigma$ to the $\Sigma(1620)$.
Therefore, the $\Lambda(1405)$ cannot be regarded as the parity partner of the $\Lambda(1116)$ and may be a flavor-singlet or exotic state.

An interesting point to mention is the ordering of the negative-parity $\Lambda$ and $\Sigma$ masses.
If one accepts the above assignment, the parity partner of the $\Lambda$, the $\Lambda(1670)$, is a little heavier than that of the $\Sigma$, the $\Sigma(1620)$.
Though $\Lambda$ and $\Sigma$ are almost degenerate in the sum rule as can be seen from Fig.~\ref{figIV}, the sum rule predicts that the $\Lambda$ is slightly heavier than $\Sigma$.
This ordering seems rather robust in the sum rule.
The fact that the SU(3) breaking effect shows up oppositely in the positive- and negative-parity states is an interesting point to be investigated further.

In the $\Xi$ channel, there exist too little experimental data to conclusively assign an observed state to the parity partner of $\Xi$.
Among the observed states, the mass of the $\Xi(1690)$~\cite{PDG} is the closest to our predicted value, 1.8 GeV. 
Therefore, the $\Xi(1690)$ can be a candidate of the parity partner, though its spin and parity are not observed yet.
However, there is another possibility that the parity partner of the $\Xi$ is not yet found because the observation of the $\Xi$ channel is relatively poor due to the experimental difficulty in the production of the particles with strangeness $-2$.
Recently, there has appeared an interesting lattice QCD calculation~\cite{Nemoto} for the negative-parity hyperon.
The predicted mass, $m_-^\Xi$, is 1.8~GeV with the statistical errors less than 0.1~GeV, which is close to the result of the present work.
This result seems to support our conjecture that the parity partner of the $\Xi$ exists around 1.8~GeV, although one should expect large systematic errors in the lattice calculation.
 
Finally, we examine how the errors of QCD parameters propagate to the results of sum rules.
The significant sources of uncertainties are the errors of $\qq$ and $m_0$, both of which are about 40 \%, while the error of $\GG$ causes rather minor modification in the results~\cite{ioffe,RRY,RRYII,BandI,SVVZ,OandP}.
In the nucleon channel, varying the $\qq$ ($m_0^2$) between this range, we find that the positive-parity masses change from 0.88 to 1.03~GeV (0.9 to 1.02~GeV) and the negative-parity from 1.44 to 1.54~GeV (1.45 to 1.56~GeV).
In the hyperon channel the errors of these parameters change the masses by $\pm0.06$~GeV on average.
The uncertainties of $m_s$, $\ss/\qq$ and $\sGs/\ss$ are about 30\%, 10\% and 40\%~\cite{ioffe,RRY,RRYII,BandI}, respectively.
These errors change the hyperon masses by $\pm0.05$~GeV on average.
The total uncertainties of the predicted masses are roughly 10\%.

\section{Summary}
We have applied the method of the QCD sum rule, in which positive- and negative-parity baryons couple with each other, to the flavor-octet hyperons and investigated the parity splittings.
We have also reexamined the nucleon in the method, which was studied in our previous paper~\cite{OUR}.
By carefully choosing the Borel weight we have obtained sum rules, which have very good Borel stability and also have a Borel window, an energy region in which the OPE converges and the pole contribution dominates over the continuum contribution for the positive- and negative-parity nucleon and hyperons.
In the $N$, $\Lambda$ and $\Sigma$ channels, the predicted masses of the positive- and negative-parity baryons reproduce the experimental ones fairly well.
The parity partners of the $\Lambda$ and $\Sigma$ are assigned to the $\Lambda(1670)$ and the $\Sigma(1620)$, respectively.
This means that the $\Lambda(1405)$ state is not the party partner of the $\Lambda$ and may be a flavor-singlet or exotic state.
In the $\Xi$ channel, there are too little experimental data to conclusively assign an observed state to the parity partner and we pointed out two possibilities.
One possibility is that the $\Xi(1690)$~\cite{PDG} is the parity partner of the $\Xi$ , whose mass is closest to our predicted value, 1.8 GeV, among the observed states of the $\Xi$ channel though the spin and parity are not determined yet.
Another possibility is that the parity partner of the $\Xi$ is not yet found because the observation of the $\Xi$ channel is relatively poor due to the experimental difficulty in the production of the particles with strangeness $-2$.
In the latter case the sum rule predicts the parity partner of the $\Xi$ exists around 1.8 GeV.
This seems to be supported by a recent lattice QCD calculation~\cite{Nemoto}, which also predicts that $m_-^\Xi$ is 1.8~GeV with a statistical error of about 0.1 GeV, though one should keep in mind large systematic errors should be taken into account.

In this paper the $\Lambda(1405)$ is not assigned to the parity partner of $\Lambda$.
In order to understand the structure of the $\Lambda(1405)$ many works have already appeared.
In the nonrelativistic quark model~\cite{Isgur} the $\Lambda(1405)$ is  interpreted as the flavor-singlet negative-parity three-quark state.
On the contrary, according to the lattice QCD calculation~\cite{Nemoto} it is difficult to identify the $\Lambda(1405)$ to be the flavor-singlet three-quark state.
There is another interesting indication that the operators consisting of the four quarks and one antiquark play an important role in the QCD sum rule approach~\cite{Liu}.
We will report on the $\Lambda(1405)$ as well as the flavor-singlet three-quark state in the present approach elsewhere.

\def\Ref#1{[\ref{#1}]}
\def\Refs#1#2{[\ref{#1},\ref{#2}]}
\def\npb#1#2#3{{Nucl. Phys.\,}{\bf B{#1}}\,(#3)\,#2}
\def\npa#1#2#3{{Nucl. Phys.\,}{\bf A{#1}}\,(#3)\,#2}
\def\np#1#2#3{{Nucl. Phys.\,}{\bf{#1}}\,(#3)\,#2}
\def\plb#1#2#3{{Phys. Lett.\,}{\bf B{#1}}\,(#3)\,#2}
\def\prl#1#2#3{{Phys. Rev. Lett.\,}{\bf{#1}}\,(#3)\,#2}
\def\prd#1#2#3{{Phys. Rev.\,}{\bf D{#1}}\,(#3)\,#2}
\def\prc#1#2#3{{Phys. Rev.\,}{\bf C{#1}}\,(#3)\,#2}
\def\prb#1#2#3{{Phys. Rev.\,}{\bf B{#1}}\,(#3)\,#2}
\def\pr#1#2#3{{Phys. Rev.\,}{\bf{#1}}\,(#3)\,#2}
\def\ap#1#2#3{{Ann. Phys.\,}{\bf{#1}}\,(#3)\,#2}
\def\prep#1#2#3{{Phys. Reports\,}{\bf{#1}}\,(#3)\,#2}
\def\rmp#1#2#3{{Rev. Mod. Phys.\,}{\bf{#1}}\,(#3)\,#2}
\def\cmp#1#2#3{{Comm. Math. Phys.\,}{\bf{#1}}\,(#3)\,#2}
\def\ptp#1#2#3{{Prog. Theor. Phys.\,}{\bf{#1}}\,(#3)\,#2}
\def\ib#1#2#3{{\it ibid.\,}{\bf{#1}}\,(#3)\,#2}
\def\zsc#1#2#3{{Z. Phys. \,}{\bf C{#1}}\,(#3)\,#2}
\def\zsa#1#2#3{{Z. Phys. \,}{\bf A{#1}}\,(#3)\,#2}
\def\intj#1#2#3{{Int. J. Mod. Phys.\,}{\bf A{#1}}\,(#3)\,#2}
\def\sjnp#1#2#3{{Sov. J. Nucl. Phys.\,}{\bf #1}\,(#3)\,#2}
\def\jtep#1#2#3{{Sov. Phys. JTEP\,}{\bf #1}\,(#3)\,#2}
\def\pan#1#2#3{{Phys. Atom. Nucl.\,}{\bf #1}\,(#3)\,#2}
\def\app#1#2#3{{Acta. Phys. Pol.\,}{\bf #1}\,(#3)\,#2}
\def\jmp#1#2#3{{J. Math. Phys.\,}{\bf {#1}}\,(#3)\,#2}
\def\cp#1#2#3{{Coll. Phen.\,}{\bf {#1}}\,(#3)\,#2}
\def\epja#1#2#3{{Eur. Phys. J.\,}{\bf A{#1}}\,(#3)\,#2}
\def\epjc#1#2#3{{Eur. Phys. J.\,}{\bf C{#1}}\,(#3)\,#2}
\def\zetf#1#2#3{{Zh. Ekps. Teor. Fiz.\,}{\bf #1}\,(#3)\,#2}
\def\yf#1#2#3{{Yad. Fiz.\,}{\bf A{#1}}\,(#3)\,#2}
\def\jpg#1#2#3{{J. Phys.\,}{\bf G{#1}}\,(#3)\,#2}

\newpage

\end{document}